\documentclass[prl,superscriptaddress,twocolumn,showkeys,preprintnumbers,floats,floatfix]{revtex4}

\usepackage{graphicx}
\usepackage{color}
\usepackage{soul}
\usepackage{amsmath}
\usepackage{setspace}

\begin{document}

\title{Single-electron shuttle based on a silicon quantum dot}

\author{K. W. Chan\footnote{Electronic mail: kokwai@unsw.edu.au}}
\affiliation{School of Electrical Engineering and Telecommunications, The University of New South Wales, Sydney 2052, Australia}
\author{M. M\"{o}tt\"{o}nen}
\affiliation{Department of Applied Physics/COMP, AALTO University, P.O. Box 14100, FI-00076 AALTO, Finland}
\affiliation{Low Temperature Laboratory, AALTO University, P.O. Box 13500, FI-00076 AALTO, Finland}
\author{A. Kemppinen}
\affiliation{Centre for Metrology and Accreditation (MIKES), P.O.
Box 9, FI-02151 Espoo, Finland}
\author{N. S. Lai}
\affiliation{School of Electrical Engineering and Telecommunications, The University of New South Wales, Sydney 2052, Australia}
\author{K. Y. Tan}
\affiliation{School of Electrical Engineering and Telecommunications, The University of New South Wales, Sydney 2052, Australia}
\affiliation{Department of Applied Physics/COMP, AALTO University, P.O. Box 14100, FI-00076 AALTO, Finland}
\author{W. H. Lim}
\affiliation{School of Electrical Engineering and Telecommunications, The University of New South Wales, Sydney 2052, Australia}
\author{A. S. Dzurak}
\affiliation{School of Electrical Engineering and Telecommunications, The University of New South Wales, Sydney 2052, Australia}

\date{\today}

\begin{abstract}
We report on single-electron shuttling experiments with a silicon
metal--oxide--semiconductor quantum dot at 300~mK.
Our system consists of an accumulated electron layer at the
Si/SiO$_{2}$ interface below an aluminum top gate with two
additional barrier gates used to deplete the electron gas locally
and to define a quantum dot. Directional single-electron shuttling
from the source to the drain lead is achieved by applying a dc
source--drain bias while driving the barrier gates with an ac voltage of frequency $f_p$.
Current plateaus at integer levels of $ef_p$ are observed up to
$f_p=240$~MHz operation frequencies.
The observed results are explained by a sequential tunneling model
which suggests that the electron gas may be heated substantially by the ac driving voltage.
\end{abstract}


\maketitle
While the Josephson voltage and the quantum-Hall resistance
standards~\cite{Jeanneret_EPJ_2009, Poirier_EPJ_2009} are routinely
used in metrology institutes worldwide, the current standard is
missing from the so-called quantum metrological triangle which
would provide a consistency check for these three quantities. The
consistency would remove any remaining doubts about the
quantum standards and justify a redefinition of the International
System of Units~\cite{Feltin_EPJ_2009}.

A metallic electron pump with a relative uncertainty of a few parts
in $10^{8}$ in its current has been achieved by utilizing seven
tunnel junctions in series, forming six gated
islands~\cite{Keller_APL_1996}. However, the operation frequency was
limited to the megahertz range because of adiabaticity requirements
for the high number of islands. These low current signals, of the
order of 1~pA, are sensitive to thermal fluctuations and hence are
not satisfactory for the planned quantum metrological triangle
experiments~\cite{Feltin_EPJ_2009}. Other single-electron pumping experiments have been
carried out in different systems such as hybrid
normal-metal--superconductor turnstiles~\cite{Pekola_NatureP_2008},
GaAs/AlGaAs nanowire quantum dots~\cite{Blumenthal_2007}, InAs nanowire double quantum
dots~\cite{Fuhrer_APL_2007}, metal--oxide--semiconductor
field-effect transistors (MOSFETs) in Si
nanowires~\cite{Fujiwara_APL_2008}, and GaAs quantum
dots~\cite{Giblin_NJP_2010}.

\begin{figure}[t]
\includegraphics[width=9.0cm]{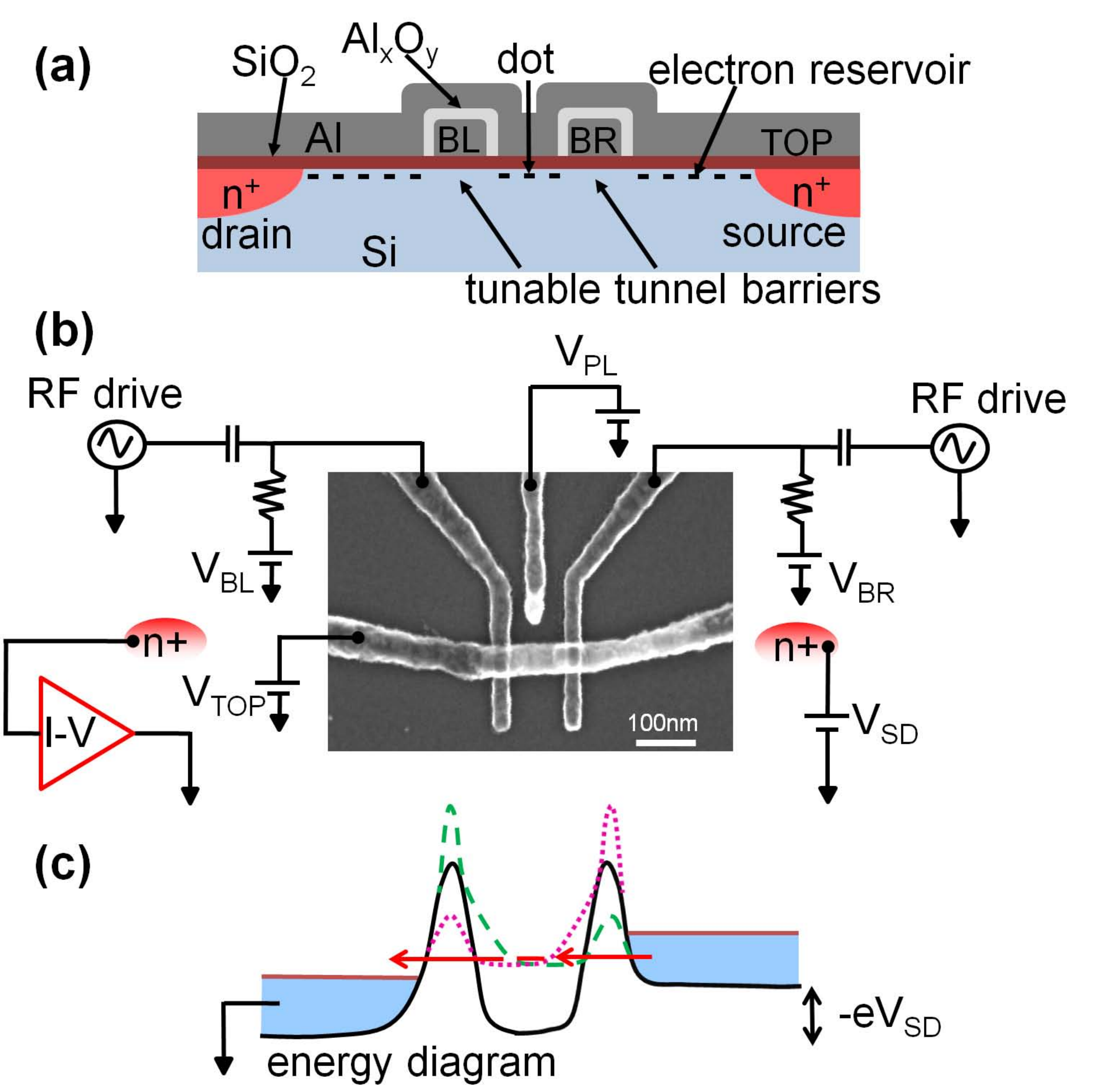}
\caption{(Color online) (a) Schematic cross section of the
fabricated silicon MOS quantum dot. Two aluminum barrier gates (BL
and BR) are below a top gate (TOP) isolated with
Al$_{x}$O$_{y}$. The source and drain are thermally diffused with
phosphorus, $n^{+}$. (b) Scanning electron microscope image of the
device with a simplified measurement setup.
(c) Energy landscape through the dot and lead reservoirs with an
illustration of the electron shuttling. When the sinusoidal ac
voltage on BR, $V_\textrm{BR}$ achieves its maximum (dashed green
line), an electron tunnels into the dot from the right electron reservoir.
After one half of an
operation period $V_\textrm{BL}$ is at its maximum value (dotted
purple line) and the electron tunnels away. Blue regions denote the
states in the leads occupied by electrons.
} \label{fig1}
\end{figure}

In this paper, we employ a silicon quantum dot~\cite{Angus_NL_2007}
shown in Fig.~\ref{fig1}(a,b) as a test-bed for the current source.
The device was fabricated on a high-resistivity ($\rho >$
10~k$\Omega$~cm at 300~K) silicon substrate. An industry-compatible
MOSFET fabrication process is adapted to realize our quantum dot
system~\cite{WHL_APL_2009}. The source and drain were thermally
diffused with phosphorus and a high-quality gate oxide was grown
thermally yielding a low Si/SiO$_{2}$ interface trap density of
2$\times$10$^{10}$~cm$^{-2}$eV$^{-1}$~\cite{McCallum_2008}. Two 30~nm wide Al barrier
gates with 90~nm separation, defined with electron beam
lithography, were subsequently oxidised on a hotplate at 150~$^{\circ}$C
for 5 minutes to form a thin 3--5~nm Al$_{x}$O$_{y}$ layer~\cite{Angus_NL_2007}.
This layer provides insulation from the overlying top gate which was
defined subsequently. An Al plunger gate was defined together with
the barrier gates with the purpose of changing the electrochemical
potential of the quantum dot.

Figure~1(a) illustrates that an electron layer is induced at the
Si/SiO$_{2}$ interface using positive top gate voltages, forming a
conduction channel between the $n^{+}$ source and drain. Two barrier
gates (BL and BR) are used to deplete the electron layer and to
define a quantum dot between them. The measurement setup is
presented in Fig.~1(b) where dc voltages were applied on all gates
and the current was measured with a room-temperature transimpedance
amplifier connected to the drain. A constant bias voltage
$V_{\textrm{SD}}$ was applied to induce a potential difference
across the device as shown in Fig.~\ref{fig1}(c). To use the device
as a single-electron shuttle, rf voltage drives were superimposed
on both barrier gate voltages. In our case, we employed two 180$^{\circ}$
out of phase sinusoidal waves, ideally keeping the electrochemical
potential of the dot fixed and moving the dot in the lateral
direction~\cite{Kouwenhoven_PRL_1991}. As shown in Fig.~\ref{fig1}(c), the dot captures an
electron when it is close to the right reservoir with the right
barrier gate voltage, $V_\textrm{BR}$, close to its maximum and the
left barrier gate voltage, $V_\textrm{BL}$, close to its minimum. To
complete a single operation cycle, an electron is ejected when the
dot is close to the left reservoir. Thus the dot is analogous to a
shuttle taking electrons one by one from the right to the left. An electrostatic
simulation using Technology Computer-Aided Design (TCAD) shows that the rf voltages applied to the barrier gates
during the shuttling operation induces dot movement in the nanometer scale. In
contrast to previously studied electron
shuttles~\cite{Gorelik_1998}, no mechanical degrees of freedom play
a role here and the frequency of the electron transport is
controlled externally.

Sufficiently low dc voltage levels at the barriers ensure that
tunneling can only take place through a tunnel junction when the ac voltage
is close to its maximum, and the high ac amplitude renders this
process very likely. The charging energy of the dot and the bias
voltage determine the number of electrons, $n$, that can tunnel
through the device during a single cycle. Ideally, the induced dc
current is given by
\begin{equation}\label{eq:chargepump}
I_\textrm{DC} = \pm nef_p,
\end{equation}
where $e$ is the electron charge and $f_{p}$ the operation
frequency.

Figure 2(a) shows the dc current through the device as a function of
the left and right barrier gate voltages with no applied ac driving voltage.
Coulomb blockade oscillations near the threshold voltage are
observed as diagonal lines. The almost horizontal
current maximum near $V_\textrm{BR}=0.56$~V indicates that there may
be a defect below the right barrier. Although we do not employ the
defect in the shuttling operation, it may lead to errors in the
current and asymmetry in the capacitances between the reservoirs and
the dot.
Figure 2(b) shows the current through the turnstile during the
shuttling operation as a function of the source--drain bias and the
plunger gate voltage. We used a peak-to-peak amplitude of $A_{\textrm{BL}} = A_{\textrm{BR}}$ = 150~mV as the ac driving voltages on the left and right barriers respectively.
Clear current plateaus at multiples of $ef_p$
are observed indicating that our qualitative shuttling principle in
Fig.~\ref{fig1}(c) is in action. The charging energy,
$E_{\textrm{C}}=e^2/(2C_{\Sigma})$, extracted from this plot equals
1.4~meV~$\approx 16$~K$\times k_B$, and thus prevents
errors due to thermal fluctuations at temperatures $\ll 16$~K
and possibly also from $1/f$ noise~\cite{Yukinori_JAP_2005}. The total capacitance of the dot is denoted by $C_\Sigma$.

\begin{figure}[t]
\includegraphics[width=9.0cm]{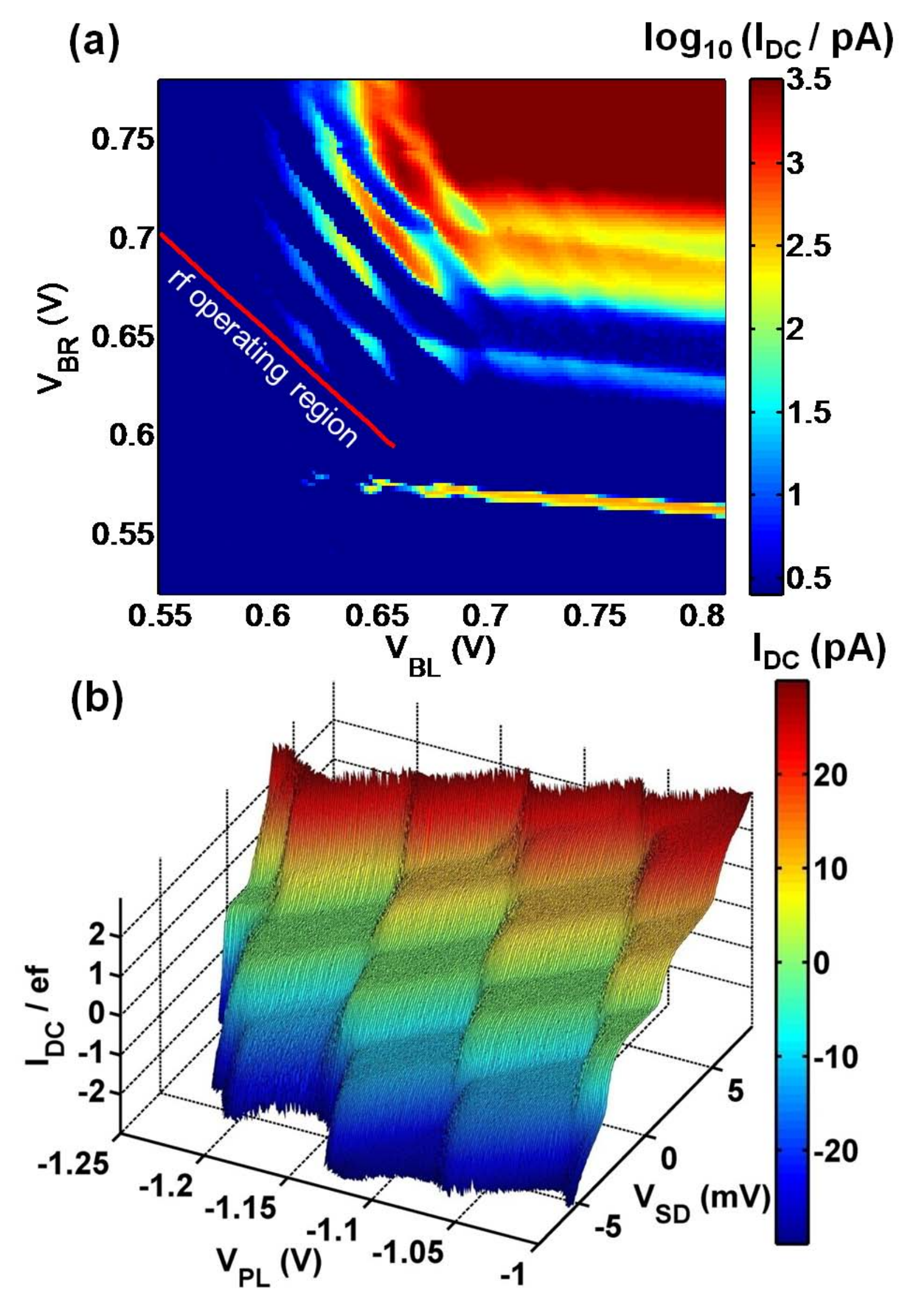}
\caption{(Color online) (a) Current through the device without the
rf drive as a function of the left and right barrier gate voltages.
Here, $V_{\textrm{SD}}$=1~mV, $V_{\textrm{TOP}}$=1.3~V, and
$V_{\textrm{PL}}$=$-$1~V. Coulomb blockade oscillations are clearly
visible near the turn-on voltages ($V_{\textrm{BL}}$=610~mV and
$V_{\textrm{BR}}$=650~mV). The red line shows the locus of ($V_{\textrm{BL}}, V_\textrm{{BR}}$) used in Fig.~3(b). (b)~Current through the device as a function of the source--drain bias and the plunger gate voltage with the rf drive applied for
$V_{\textrm{TOP}}$=1.3~V, $V_{\textrm{PL}}$=$-$1~V, $\langle
V_{\textrm{BL}}\rangle$=598~mV, and $\langle V_{\textrm{BR}}
\rangle$=654~mV. The shuttling operation frequency is
$f_{\textrm{p}}$=60~MHz and the peak-to-peak amplitude of the
sinusoidal rf drive is $A_{\textrm{BL}}$=150~mV and
$A_{\textrm{BR}}$=150~mV for the left and right barrier,
respectively.} \label{fig2}
\end{figure}

Figure~3(a) shows current plateaus at different frequencies in
agreement with Eq.~(\ref{eq:chargepump}) up to 240~MHz. Figure~3(b)
shows the shuttle current with parameters tuned to optimize the
flatness of the plateaus at currents near $ef_p$ and zero.
At the plateau near zero current shown in Fig.~3(c), the mean
current is 9~fA for $V_{\textrm{SD}}$ ranging from $-$0.5~mV to 0.5~mV.
The origin of this finite current remains unclear.

\begin{figure}[t]
\includegraphics[width=9.0cm]{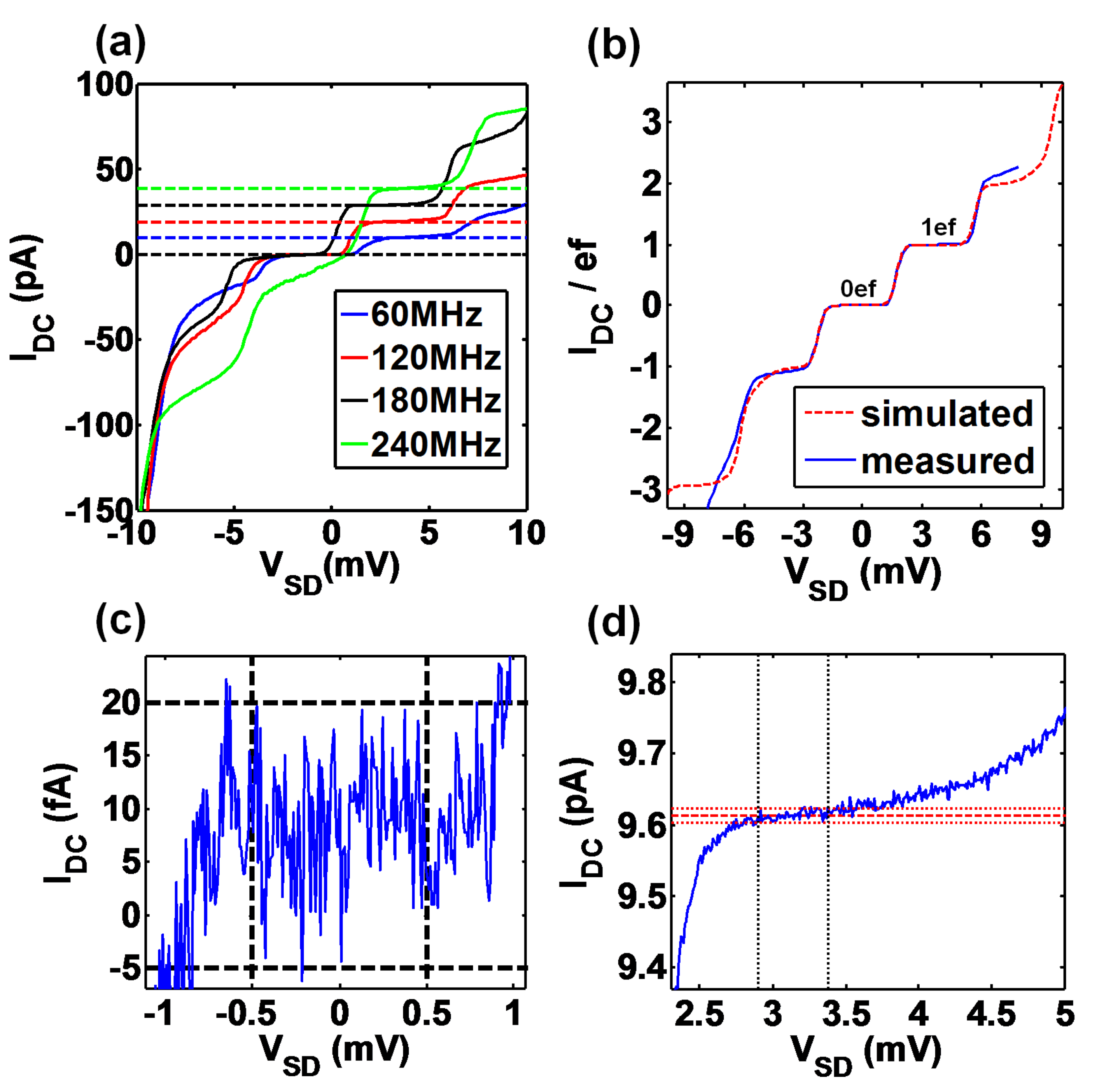}
\caption{(Color online) (a) Charge shuttling current measured for
$f_{\textrm{p}}$=60, 120, 180 and 240~MHz with other
parameters identical to those in Fig.~\ref{fig2}(b). The dashed lines show the
current values at multiples of $e\times 60$~MHz. (b) Measured charge
shuttling current (solid blue line) for $f_{\textrm{p}}$=60~MHz,
$A_{\textrm{BL}}$=$A_{\textrm{BR}}$=120~mV,
$V_{\textrm{TOP}}$=1.271~V, $V_{\textrm{PL}}$=$-$1.13~V, $\langle
V_{\textrm{BL}}\rangle$=598~mV, and $\langle
V_{\textrm{BR}}\rangle$=654~mV. The trace of the control voltages
$(V_{\textrm{BL}},V_{\textrm{BR}})$ is shown as the red line in
Fig~2(a). The measured current is compared with the simulation
(dotted red line) based on a sequential tunneling model with
variable tunneling resistances in the barriers. (c,d) Measured
current with the same settings as in panel (b) for (c) the
zero-current $(n = 0)$ and (d) $ef_p$ $(n = 1)$ plateaus. \label{fig3}}
\end{figure}

The $ef_p$ plateau for $n = 1$ is shown in Fig. 3(d). In contrast to the
zero-current plateau, the current near $ef_p$ does not achieve its
lowest gradient at the center of the plateau but at somewhat lower
voltages. Typically for semiconductor single-electron sources,
the quantized current is measured at this low-gradient
point~\cite{Blumenthal_2007, Giblin_NJP_2010}. This low-gradient
point is estimated to be at bias voltages ranging from 2.9~mV to
3.4~mV where we have 29 data points with the mean value 9.6138~pA.
This mean value is very close to the ideal current $e\times
60\textrm{ MHz}=9.6131$~pA. However, the gain of the transimpedance
amplifier was calibrated with a 1~G$\Omega$ resistor up to 2~nA
current with an estimated error of $10^{-4}$. This
calibration yields only 2\% accuracy for 10~pA current, and hence we
cannot claim the absolute accuracy of the device to be higher than this level. In
practice, the calibration is expected to be more accurate since the
gain of the amplifier is highly linear and the dominating error in
the calibration is likely to arise from the 1~G$\Omega$ resistor.
The horizontal dotted lines denoting $10^{-3}$  error in
Fig.~\ref{fig3}(d) illustrate the flatness of the plateau. Even a
more pessimistic estimate of the relative error of $2\times 10^{-3}$
can be given based on the value of the current not measured at the
low-gradient region, but in the middle point of the plateau at
$V_\textrm{SD}\approx 3.8$~mV. The argument for this estimate is
that there may exist unknown error processes which tend to decrease the
current in their part with increasing bias voltage and hence artificially lower
the gradient for a certain bias region. A thorough theoretical
analysis and modeling of the different tunneling processes is
required to make a definite conclusion whether this is possible.

Figure~\ref{fig3}(b) also shows a comparison of the measured current
to a simulation result using a sequential
tunneling model. The model has been utilized previously for metallic
systems~\cite{Pekola_NatureP_2008}, but in our case the tunneling
resistances of the left and right barrier were taken to depend
exponentially on the respective gate voltage. This dependence was
chosen to agree with the measurement in Fig.~\ref{fig2}(a). The
capacitance values were primarily extracted from dc measurements but
a slight fine tuning was carried out to fit well with the
experimental data due to issues such as the difference in the attenuation in the rf lines.
In addition, a shift of 0.1~mV was applied to the
source--drain bias justified by the non-vanishing input voltage of
the transimpedance amplifier. It turned out that the sequential
tunneling model did not agree with the measurements if the
temperature of the cryostat thermometer, 300~mK, was used. Instead,
an electron temperature of 1.5~K gave the good fit shown in
Fig.~\ref{fig3}(b). In the case of no rf driving, a good agreement with the measurements of the current--voltage characteristics near Coulomb blockade was obtained if the temperature was set to 300~mK in the simulation (data not shown). This suggests that the source and drain electron
reservoirs may heat up due to the rf drive of the barrier gates.
Thus the high charging energy of the dot is essential to prevent
errors from the thermal fluctuations. In future, we aim to improve
our model to be able to compare the shuttle current up to high
accuracies at the plateaus shown in Fig.~\ref{fig3}(c,d).

Our studies show also that the asymmetry in the capacitance from the
dot to the left and right reservoirs, or the asymmetry in the amplitudes
of the sinusoidal gate drives, can cause the electrochemical
potential of the dot to oscillate during the control cycle. This in
turn has the effect that the current plateaus at negative and
positive bias are of different quality. If the asymmetry is large
enough, a plateau at $I = ef_p$ can appear at zero or even negative bias.
To compensate for this uncontrolled asymmetry, we suggest the use of a
third rf driving voltage on the plunger gate and an improved
coupling to the plunger gate~\cite{WHL_APL_2009}.

In summary, the single-electron shuttle is a
promising current source with demonstrated current plateaus at a few
tens of pA. Although these results are still far from the
requirements of a metrological current standard, sample and pulse
sequence optimization has the potential to improve the accuracy and
yield to the desired level.

The authors thank D.~Barber, R.~P.~Starrett, and J.~Szymanska for
technical support, and J.~P.~Pekola for discussions and cryostat
time. We acknowledge Academy of Finland, Emil Aaltonen Foundation,
and Technology Industries of Finland Centennial Foundation for
financial support. This research was also supported in part by the Australian National Fabrication Facility,
the Australian Research Council Centre of Excellence for Quantum Computation and
Communication Technology (project number CE110001029), and by the
U.\ S.\ National Security Agency and U.\ S.\ Army Research Office
(under Contract No. W911NF-08-1-0527).

\end{document}